# A Systematic Review of Passive Cooling Strategies Integrating Traditional Wisdom and Modern Innovations for Sustainable Development in Arid Urban Environments

Shiva Manshour*, Steffen Lehmann

University of Nevada, Las Vegas, Las Vegas, Nevada, 89154, USA
shiva.manshour@unlv.edu, steffen.lehmann@unlv.edu

## Abstract

Urban environments in hot-arid regions are increasingly challenged by rising temperatures, rapid urban expansion, and the heavy reliance on energy-intensive mechanical cooling systems. This study presents a systematic review of peer-reviewed literature from 1980 to 2025 to assess both traditional and contemporary passive cooling strategies tailored for arid urban settings. Following PRISMA 2020 guidelines, 30 high-quality studies were selected from leading databases including Scopus, Web of Science, ScienceDirect, and Google Scholar. The selected works span diverse geographical contexts—from the Middle East and North Africa to parts of South Asia—and apply a range of methods including field experiments, computer-based simulations, and qualitative analyses. Findings highlight strong consensus around core passive principles such as solar control, natural ventilation, and the use of thermal mass. Vernacular solutions like courtyards, wind towers, and thick masonry walls remain effective, while innovations such as cool roofs, phase change materials, and parametric optimization techniques expand the design toolkit. Nevertheless, the implementation of these strategies is often limited by climate variability, cultural shifts, regulatory frameworks, and economic feasibility. The review concludes that context-sensitive, hybrid solutions—combining traditional knowledge with modern technology—hold the greatest potential for achieving sustainable thermal comfort. To be effective, these approaches must be supported by climate-adaptive urban planning, user-centered design, and updated building regulations. The study offers practical insights for architects, planners, and policymakers aiming to create resilient, low-carbon cities that harmonize cultural identity with environmental responsibility.

## 1. Introduction

Arid urban regions are increasingly threatened by extreme climatic conditions—characterized by high solar radiation, wide diurnal temperature variations, minimal precipitation, and low humidity—conditions that severely impact human health, environmental sustainability, and urban livability (Salameh & Touqan, 2022; Song et al., 2021; Al-Tamimi, 2022; Taheri, 2015; Bahrpeima & Sargazi, 2022). These impacts are further intensified by climate change and the urban heat island (UHI) effect, which elevate cooling demands and energy consumption, with mechanical systems often accounting for over 60% of total energy use in buildings (Pathan, 2023; Jega & Al-Din, 2023; Fereidani et al., 2021; Ibrahim et al., 2021).

In response, passive cooling strategies—defined as design approaches that maintain thermal comfort through natural means—offer sustainable alternatives that significantly reduce energy dependence and greenhouse gas emissions (Beigli & Lenci, 2016; Gervais, 2012; Matallah et al., 2023; Athmani et al., 2023). Rooted in vernacular traditions, these strategies include architectural features such as courtyards, wind catchers, high thermal mass construction, and water-integrated microclimates, all of which demonstrate high efficacy in mitigating urban heat (Izadpanahi et al., 2021; Eltrapolsi, 2016; Chohan & Awad, 2022; Elian, 2024).

However, modern urban development frequently neglects these proven methods, favoring energy-intensive designs that are climatically insensitive and environmentally detrimental (Mazhar, 2024; Salameh & Touqan, 2023; Ibrahim et al., 2021; Lotfata et al., 2024). Although empirical studies affirm the performance of passive strategies in reducing indoor and outdoor temperatures (Al-Tamimi, 2022; Song et al., 2021; Gomaa et al., 2024), a research gap persists regarding the integration and comparative evaluation of traditional and modern cooling methods. Bridging this gap is crucial for developing resilient urban solutions aligned with both heritage and innovation (Fereidani et al., 2021; Necira et al., 2024).

Accordingly, this study conducts a systematic review of passive cooling strategies in arid cities, aiming to synthesize historical knowledge and emerging technologies from 2000 to 2025. Three guiding research questions structure the inquiry:

1. **Which traditional methods are most effective for passive cooling?** Evidence supports the effectiveness of courtyards, wind towers, thick thermal mass walls, water features, and narrow streets in moderating microclimates (Izadpanahi et al., 2021; Song et al., 2021; Chohan & Awad, 2022; Beigli & Lenci, 2016; Elian, 2024; Salameh & Touqan, 2022; Al-Tamimi, 2022).

2. **How have modern strategies improved or replaced traditional methods?** Contemporary innovations include high-performance insulation, reflective coatings, phase change materials (PCM), and climate-optimized geometries (Fereidani et al., 2021; Tarek et al., 2024; Necira et al., 2024; Song et al., 2021; Elnabawi et al., 2024; Athmani et al., 2023; Mohammed & Radha, 2022).

3. **What indicators measure the success of passive cooling strategies?** Key metrics include thermal comfort indices (e.g., PMV), reductions in cooling load and indoor temperature, energy and emissions savings, and mitigation of UHI effects. Indirect outcomes such as improved public health and outdoor habitability also serve as success criteria (Mushtaha et al., 2021; Khechiba et al., 2023; Benbrahim et al., 2025; Lotfata et al., 2024).

Following PRISMA guidelines, this review draws from databases such as Scopus, Web of Science, ScienceDirect, and Google Scholar, analyzing literature using NVivo for qualitative synthesis. Inter-rater reliability was maintained through dual screening and resolution procedures. Through this methodological rigor, the review aims to offer architects, urban planners, and policymakers actionable insights for integrating traditional wisdom with modern innovations—contributing to the creation of sustainable, climate-adaptive urban environments.

## 2. Theoretical Background
### 2.1 Passive Cooling

Passive cooling encompasses architectural and environmental design strategies aimed at reducing indoor heat gain and enhancing thermal comfort without reliance on mechanical systems (Song et al., 2021; Ibtissame, Imane, & Salima, 2022). Rooted in thermodynamic principles and building physics, passive techniques manipulate environmental forces—solar radiation, wind, and thermal mass—to minimize internal temperature fluctuations, making them especially relevant in arid climates (Beigli & Lenci, 2016; Salameh & Touqan, 2022). The theoretical framework classifies passive cooling into three primary categories: heat gain prevention, natural ventilation, and thermal inertia (Song et al., 2021). Heat gain prevention employs shading devices, reflective materials, and insulation to block solar radiation (Al-Tamimi, 2022; Athmani et al., 2023), while natural ventilation utilizes wind-induced and buoyancy-driven airflow for heat removal (Chohan & Awad, 2022; Ayoobi, Ekimci, & Inceoğlu, 2024). Thermal inertia refers to the capacity of high-mass materials, such as adobe and stone, to absorb and slowly release heat, delaying indoor temperature peaks (Izadpanahi, Farahani, & Nikpey, 2021; Eltrapolsi, 2016). Complementing these are evaporative cooling and radiative cooling strategies. Courtyard fountains and vegetation facilitate cooling through evaporation, while roof coatings and nocturnal sky exposure enhance radiative heat dissipation (Necira et al., 2024; Elian, n.d.; Shashua-Bar, Pearlmutter, & Erell, 2009; Ochoa, Marincic, & Coch, 2022). Table 1 provides a summary of these categories and mechanisms.

Recent scholarship further enriches the theoretical landscape of passive cooling by addressing a range of scales and typologies. For instance, research on urban renewal underscores the necessity of integrating passive strategies in the rehabilitation of traditional neighborhoods to support long-term sustainability (Akalp & Ayçam, 2024). Bibliometric analyses have mapped the evolution of global passive cooling research, highlighting key theories and emerging methods (Bema & Ozarisoy, 2024). Reviews also emphasize the significance of large-scale urban initiatives, such as city-wide cooling interventions and policy frameworks aimed at mitigating heat stress (CAIRO, 2023). The use of advanced materials and super-cool technologies has demonstrated promising outcomes in educational and public buildings (Correia et al., 2024). Microclimatic adaptation is further exemplified by studies investigating the role of spatial configuration and landscape elements in creating comfortable outdoor environments (Elian, n.d.). Recent systematic reviews explore the risks of overheating in passive house designs, underlining the importance of dynamic modeling and post-occupancy evaluation (Farrokhirad et al., 2024; Nagasue et al., 2024). Nature-based solutions and green infrastructure are increasingly recognized for their capacity to reduce urban heat and promote low-carbon cities (Qingchang, 2023). Moreover, integrating passive and active design scenarios in hot, dry climates has proven effective in achieving positive energy buildings (Sarir & Sharifzadeh, 2024), while the impact of hybrid ventilation strategies in arid regions continues to be a focus of contemporary analysis (Sheikha, 2024).

**Table 1 Classification of passive cooling strategies (adapted from Song et al., 2021; Beigli & Lenci, 2016; Al-Tamimi, 2022).**

| Category | Strategy Examples | Cooling Mechanism |
|---|---|---|
| Heat Gain Prevention | External shading, reflective surfaces, insulation | Limits solar radiation and conductive heat |
| Natural Ventilation | Wind towers, cross ventilation, stack effect | Enhances airflow and heat removal |
| Thermal Inertia & Storage | High-mass walls, night flushing, earth coupling | Stores daytime heat, releases at night |
| Evaporative Cooling | Fountains, water bodies, green roofs | Lowers air temperature via evaporation |
| Radiative Cooling | Roof coatings, nighttime sky exposure | Releases heat through infrared radiation |

Empirical studies validate the effectiveness of these approaches. For instance, vegetation-integrated urban landscapes have been shown to reduce ambient air temperatures by up to 4°C (Shashua-Bar et al., 2009), while oasis landscapes demonstrate the value of natural groves for cooling in desert environments (Matallah et al., 2023). Recent developments also explore **hybrid systems**, combining passive and active methods such as smart facades and sensor-driven controls to optimize comfort while minimizing energy use (Mushtaha et al., 2021; Fereidani, Rodrigues, & Gaspar, 2021). Despite technological advances, literature emphasizes that passive-first approaches—rooted in local context and material intelligence—remain central to sustainable design. Moreover, performance-based simulation tools, such as CFD models and thermal comfort analysis, have transformed passive cooling from intuitive practice into a scientifically optimized process (Dhanraj, Vora, & Naik, 2023; Tarek, Aly, & Ragab, 2024). Effective implementation requires a holistic design approach integrating climate, geometry, material selection, and occupant behavior. In summary, passive cooling represents both a historical precedent and a contemporary imperative for arid urban environments. Its theoretical underpinnings support resilient, low-energy design, and its fusion with advanced technologies provides a robust framework for climate-responsive architecture (Sharami & Hosseini, 2024; Salameh & Touqan, 2022).

## 2.2 Historical Background and Traditional Wisdom

Traditional architecture in arid regions of the Middle East and North Africa demonstrates a sophisticated understanding of passive cooling, developed over centuries through empirical adaptation to climate (Salameh & Touqan, 2022). These vernacular systems utilized indigenous materials and spatial configurations to mitigate solar heat gain and enhance ventilation—principles that remain highly relevant for contemporary sustainable design.

- **Courtyards and Orientation:** Central courtyards—ubiquitous in Islamic and Persian architecture—served as thermal regulators, providing shaded microclimates and enabling nocturnal radiative cooling (Beigli & Lenci, 2016; Elian, n.d.). High surrounding walls and internal water features (e.g., hawdz pools) facilitated convective and evaporative cooling, with studies reporting temperature reductions of several degrees (Sharami & Hosseini, 2024; Izadpanahi, Farahani, & Nikpey, 2021). Courtyard dimensions were carefully proportioned to maximize shade and airflow, often creating a stack effect that enhanced single-sided ventilation, particularly in inward-facing housing typologies (Chohan & Awad, 2022; Sharami & Hosseini, 2024).

- **Wind Catchers (Badgir) and Ventilation Towers:** Wind catchers, or *badgir*, were integral to buildings across Iran and the Persian Gulf, channeling prevailing winds into living spaces and sometimes incorporating subterranean pools or qanats for additional evaporative cooling (Chohan & Awad, 2022; Sharami & Hosseini, 2024). These towers operated via pressure differentials and buoyancy, and some functioned even without wind due to nocturnal radiative cooling. Studies confirm their effectiveness in lowering indoor temperatures by up to 10 °C (Elian, n.d.; Salameh & Touqan, 2022).
- **Mashrabiya and Shading Devices:** The *mashrabiya*—a carved wooden lattice enclosing windows or balconies—offered combined benefits of privacy, shading, and evaporative cooling. Originally used to store water jars, mashrabiyas humidified and cooled incoming air while significantly reducing solar radiation and glare (Salameh & Touqan, 2022). Similar perforated screens, such as *shanasheel* and *jali*, exemplify culturally embedded passive cooling solutions integrating aesthetics and environmental performance.
- **Thermal Mass and Earth Coupling:** Traditional buildings often featured thick adobe or stone walls with high thermal mass, which absorbed solar heat during the day and released it at night, reducing indoor temperature fluctuations (UN-Habitat, 2021; Sharami & Hosseini, 2024). Light-colored exteriors reflected solar radiation, while underground spaces (e.g., *Hozkhaneh* rooms or *yakhchal* ice storage chambers) utilized earth coupling to maintain stable indoor temperatures well below outdoor levels.
- **Narrow Shaded Alleys and Urban Form:** Urban layouts with compact, narrow alleyways provided extensive self-shading and harnessed prevailing winds for ventilation. These microclimatic strategies were seen in medinas across North Africa, as well as in traditional cities like Yazd and Kerman (Chohan & Awad, 2022). Orientation and street geometry were critical in enhancing pedestrian comfort and passive airflow. The design of public spaces in hot-arid urban environments must integrate ecological sensitivity and cultural relevance to enhance livability and long-term urban resilience (Manshour, 2025). Table 2 summarizes the major traditional passive cooling elements identified across arid climatic zones:

**Table 2 Traditional Passive Cooling Strategies in Hot-Arid Climates (adapted from Sharami & Hosseini, 2024; Chohan & Awad, 2022; Salameh & Touqan, 2022).**

| Strategy | Cooling Mechanism | Examples (Regions) | Sources |
| --- | --- | --- | --- |
| Central Courtyard | Shading, stack ventilation, radiative and evaporative cooling | Iran (Kashan), Maghreb, Arabian Peninsula | Elian, n.d.; Sharami & Hosseini, 2024 |
| Wind Catcher Tower (*Badgir*) | Wind-driven ventilation, evaporation, stack effect | Yazd, Cairo, Dubai, Bahrain | Sharami & Hosseini, 2024; Chohan & Awad, 2022 |
| Mashrabiya / Lattice Screen | Airflow shading, evaporative inlet cooling | Cairo, Damascus, Iraq, Morocco | Salameh & Touqan, 2022 |
| Thick Mudbrick/Stone Walls | Thermal mass, delayed heat transfer, solar reflectivity | Shibam (Yemen), Siwa (Egypt), Saharan ksars | UN-Habitat, 2021; Sharami & Hosseini, 2024 |
| Earth-Sheltered Rooms | Earth coupling, stable subsurface temperatures | Yazd, Matmata (Tunisia), Afghanistan | Sharami & Hosseini, 2024 |
| Narrow Shaded Alleys | Self-shading, microclimate enhancement, wind alignment | Fez, Tunis, Masdar City (Abu Dhabi), Persian historic towns | Chohan & Awad, 2022 |

In conclusion, traditional architecture in arid zones exhibits a comprehensive and environmentally attuned approach to passive cooling. These techniques—developed through localized knowledge and resource-conscious practices—offer valuable precedents for modern sustainable design and urban resilience strategies.

**2.3 Modern Innovations in Passive Cooling Design**

Contemporary architecture has expanded the passive cooling paradigm by introducing advanced materials, systems, and digital design tools that enhance or adapt traditional strategies for present-day building needs. These innovations aim to improve thermal performance, energy efficiency, and comfort, particularly in arid climates where cooling demand is high (Chohan & Awad, 2022; Al-Tamimi, 2022). Key advancements can be grouped into material enhancements, hybrid systems, adaptive facades, and computational optimization.

- **High-Reflectivity Roofs and Surface Treatments:** Cool roofs—engineered with reflective coatings and high emissivity materials—reduce surface temperatures by 20–40 °C, limiting solar heat gain and decreasing cooling loads (Elnabawi et al., 2024; Ibrahim et al., 2021; Necira et al., 2024). Modern variants surpass traditional whitewashing in durability and spectral performance. Urban heat island mitigation initiatives increasingly incorporate these materials, which can cut annual energy use by up to 23% (Benbrahim et al., 2025). Green roofs provide additional evaporative cooling and insulation (Ibtissame et al., 2022; Salameh & Touqan, 2023).
- **Phase Change Materials (PCMs):** PCMs offer latent heat storage by absorbing and releasing heat during phase transitions, allowing slim profiles to replicate the thermal damping of massive walls (Fereidani et al., 2021; Pathan, 2023). PCM-embedded products can maintain indoor temperatures near comfort thresholds and reduce peak temperatures by 2–5 °C, achieving energy savings of 15–30% (Dhanraj et al., 2023; Athmani et al., 2023). Recent developments include nano-encapsulation and bio-based PCMs, enabling integration into ceilings, walls, and furniture (Izadpanahi et al., 2021).
- **Hybrid Ventilation and Mixed-Mode Systems:** Blending natural and mechanical ventilation, hybrid systems use sensors, motorized vents, and low-energy fans to automate airflow responses to outdoor conditions (Song et al., 2021; Al-Tamimi, 2022). Modern wind towers in the Persian Gulf region, such as those in Masdar City, combine downdraft cooling with solar-powered fans and mist systems to achieve ambient reductions of 5–7 °C (Mazhar, 2024; Mohammed & Radha, 2022). Night flushing and demand-controlled ventilation extend the passive potential of these systems while ensuring performance in dense urban environments (Beigli & Lenci, 2016; Ibrahim et al., 2021).
- **Smart Shading and Responsive Facades:** Dynamic facades, inspired by elements such as the *mashrabiya*, adapt in real-time to sun angle and intensity. Examples include the kinetic facade of Al Bahar Towers, which adjusts to solar exposure and reduces cooling demand by up to 50% (Chohan & Awad, 2022; Sharami & Hosseini, 2024). Other innovations include electrochromic glazing, PV-integrated screens, and biomimetic envelope designs. These systems actively regulate solar gain and daylighting, optimizing comfort while preserving environmental responsiveness (Fereidani et al., 2021; Mazhar, 2024).

- **Computational Design and Parametric Optimization:** Parametric modeling and simulation tools (e.g., Grasshopper with Ladybug/Honeybee) allow designers to iteratively evaluate passive strategies under climatic conditions, significantly enhancing design precision (Dhanraj et al., 2023; Gervais, 2012). Case studies such as the King Hussein Mosque demonstrate how optimized courtyard geometry and fenestration improve thermal comfort and reduce energy intensity (Pathan, 2023; Mohammed & Radha, 2022). CFD and energy modeling tools enable real-time feedback, replacing rule-of-thumb methods with data-driven performance tuning (Lotfata et al., 2024).
- **Emerging Materials: Radiative Cooling Coatings:** Passive Daytime Radiative Cooling (PDRC) materials represent a novel frontier, using photonic coatings to reflect solar radiation and emit infrared heat to the sky even under direct sunlight (Mehmood et al., 2022; Dhanraj et al., 2023). Early prototypes have maintained surface temperatures up to 8 °C below ambient and are being tested as adjuncts to HVAC systems.

Table 3 Summary of Modern Enhancements in Passive Cooling Strategies

| Innovation | Cooling Mechanism | Performance Highlights | Sources |
|---|---|---|---|
| Cool Roofs and Reflective Surfaces | Radiative heat rejection | 10–23% energy reduction; surface cooling 20–40 °C | Elnabawi et al., 2024; Benbrahim et al., 2025 |
| Phase Change Materials (PCMs) | Latent thermal storage | 15–30% cooling energy reduction; temperature damping by 2–5 °C | Pathan, 2023; Athmani et al., 2023 |
| Hybrid Ventilation Systems | Combined passive/mechanical ventilation | ~23% energy savings; improved comfort via automated controls | Mohammed & Radha, 2022; Beigli & Lenci, 2016 |
| Smart Facades (e.g., Al Bahar) | Adaptive solar control | ~50% cooling load reduction; glare mitigation | Mazhar, 2024; Chohan & Awad, 2022 |
| Green Roofs/Walls | Evaporative and shading cooling | ~5 °C indoor temp reduction; reduced urban heat island effect | Ibtissame et al., 2022; Athmani et al., 2022 |
| Parametric and CFD Simulation | Climate-specific optimization of form and envelope | Up to 11% EUI savings; ~15% improvement in thermal comfort | Mohammed & Radha, 2022; Taghavi Araghi, 2022 |
| PDRC Coatings (Emerging) | Daytime sky-directed radiative cooling | Surface temps up to 8 °C below ambient during midday sun | Mehmood et al., 2022; Dhanraj et al., 2023 |

Modern innovations in passive cooling do not replace traditional approaches but rather extend their principles through technology and performance-driven design. By combining vernacular logic with contemporary capabilities, designers can achieve high-impact thermal comfort solutions that align with the goals of sustainable urban development (Beigli & Lenci, 2016; Mazhar, 2024).

To provide a clearer understanding of how passive cooling approaches have evolved over time, Table 4 presents a comparative summary of traditional and modern passive cooling strategies. This table synthesizes key elements from the reviewed literature, highlighting the core mechanisms, contextual relevance, materials, and performance features of each approach. It illustrates how historical vernacular solutions grounded in environmental adaptation have informed, and in many cases been enhanced by, contemporary innovations in material science, digital design, and hybrid systems. The aim is to underscore both the continuity and transformation in passive cooling practices and to reveal opportunities for integrated, climate-resilient design in arid urban contexts.

**Table 4 Comparative Summary of Traditional vs. Modern Passive Cooling Strategies (Adapted from Sharami & Hosseini, 2024; Chohan & Awad, 2022; Salameh & Touqan, 2022; Elnabawi et al., 2024; Pathan, 2023; Dhanraj et al., 2023)**

| Category | Traditional Element | Modern Equivalent / Innovation | Cooling Mechanism |
|---|---|---|---|
| **Shading** | Courtyards, Mashrabiya, Shanasheel, narrow alleyways | Dynamic facades (e.g., Al Bahar Towers), PV-integrated louvers | Solar radiation control, shading, and daylight regulation |
| **Ventilation** | Wind catchers (Badgir), stack ventilation, open courtyards | Hybrid ventilation systems with sensors, solar chimneys | Natural airflow, buoyancy-driven and wind-induced cooling |
| **Thermal Mass / Insulation** | Thick adobe/stone walls, earthen materials | Phase Change Materials (PCMs), nano-insulation layers | Thermal storage and delayed heat transfer |
| **Evaporative Cooling** | Courtyard fountains, hawdz pools, qanats, vegetation | Green roofs, mist systems, water walls | Evaporation-induced temperature reduction |
| **Radiative Cooling** | Night sky exposure via open courtyards and roof terraces | Passive Daytime Radiative Cooling (PDRC) coatings | Heat dissipation via long-wave infrared radiation |
| **Material Reflectance** | Light-colored mud plaster, whitewashed domes | Cool roofs with reflective membranes, emissive surface treatments | Minimized solar absorption, roof surface temperature drop |
| **Design Tools / Optimization** | Proportional design by empirical knowledge | Parametric modeling, CFD, thermal simulations (e.g., Ladybug, Envi-met) | Performance-driven, site-specific design refinement |

As shown in Table 4, traditional strategies such as courtyards, wind towers, and thick thermal mass walls rely heavily on passive interactions with the local climate, materials, and spatial forms, while modern innovations introduce higher precision and scalability through smart technologies, advanced coatings, and parametric modeling. Importantly, these categories are not mutually exclusive; instead, they offer complementary strengths that can be selectively combined. For instance, modern materials like Phase Change Materials (PCMs) can supplement traditional adobe walls to enhance thermal inertia, or kinetic shading systems may be inspired by the logic of the mashrabiya while offering greater flexibility. Ultimately, the integration of both traditional wisdom and modern engineering allows for adaptable, culturally sensitive, and high-performance cooling solutions tailored to the diverse challenges of today's hot-arid urban environments.

## 3. Methodology

This research adopted a systematic review methodology aligned with the PRISMA 2020 framework, ensuring transparency, reproducibility, and methodological rigor. The objective was to comprehensively synthesize findings related to passive cooling in hot-arid urban contexts, incorporating both traditional and modern approaches. The review focused on peer-reviewed literature published between 2000 and early 2025, sourced from four major academic databases: Scopus, Web of Science, ScienceDirect, and Google Scholar. The selected 25-year window reflects both technological advancement and the evolution of sustainable urban design practices. Search queries were formulated using Boolean operators and included terms such as "passive cooling," "hot-arid climate," "urban environments," and "vernacular architecture." Each database's syntax was adapted accordingly, and reference management software was employed to identify and remove duplicates.

In total, 743 records were initially retrieved. After removing 123 duplicates, 620 unique records remained and were screened by title and abstract. At this stage, 510 studies were excluded for not meeting basic relevance criteria. The remaining 110 articles underwent full-text review to assess eligibility. Based on predefined inclusion criteria—namely, focus on at least one passive cooling strategy, urban setting in a hot-arid or dry climate, English-language publication, and appearance in peer-reviewed sources—80 full-text articles were excluded. These were disqualified for the following reasons: 28 did not involve relevant climate conditions, 22 lacked urban scale relevance, 20 lacked a passive cooling focus, and 10 were excluded for other methodological or contextual reasons.

Ultimately, 30 high-quality studies were selected for qualitative synthesis. Key information was extracted from each study, including location, climatic zone, passive cooling technique, analytical methods, and key findings. Data were compiled into a structured matrix using Microsoft Excel and then imported into NVivo 12 for qualitative content analysis. A thematic coding framework was developed to categorize strategies into traditional or modern typologies, as well as to identify research outcomes and implementation challenges. To ensure analytical reliability, dual coding was initially applied to a subset of studies, and a high level of inter-rater agreement was achieved. The remaining studies were coded consistently by a single researcher following the validated scheme. The final set of 30 studies spans a range of geographic regions and methodological designs, from field measurements to simulation models and comparative case studies, covering building- to city-scale interventions. This systematic evidence base forms the foundation for the results and discussion in subsequent sections of the paper.

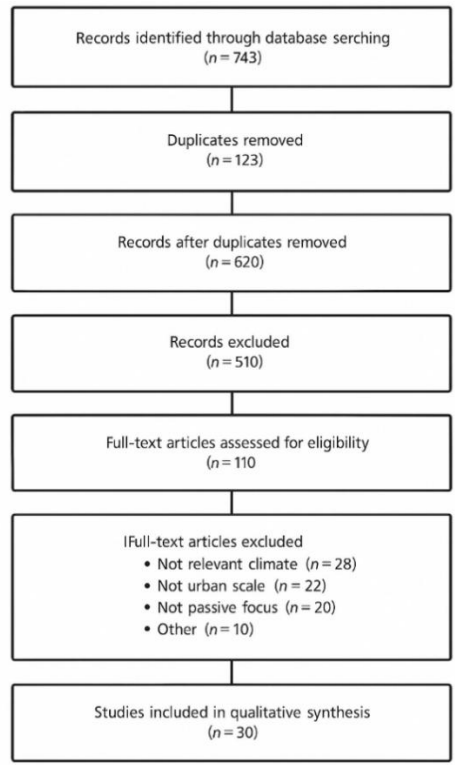

*Figure 1 PRISMA flow diagram summarizing the identification, screening, eligibility, and inclusion of studies (final n = 30 included articles). The diagram outlines the number of records at each step and the reasons for exclusions, following PRISMA 2020 guidelines*

To synthesize the diverse body of literature on passive cooling in hot-arid climates, the following table provides a concise summary of the key findings from 30 rigorously selected studies published between 1978 and 2025. Rather than listing each entry individually, this structured synthesis categorizes the studies by cooling mechanism and highlights representative works under each theme. The summarized format facilitates quick comparison of methodological approaches, performance outcomes, and recurring passive strategies across different geographic, cultural, and architectural contexts. The goal is to capture the most actionable insights that inform sustainable cooling practices in contemporary urban development.

Table 5 selected Studies and key Findings

| Theme | Key Studies | Findings |
|---|---|---|
| Courtyards & Compact Urban Form | Safarzadeh & Bahadori (2005); Soflaei et al. (2016); Izadpanahi et al. (2021); Salameh & Touqan (2023) | Courtyards reduce indoor temperature (up to 3–5°C), enhance air circulation, and serve as effective microclimate regulators in hot-arid climates. |
| Windcatchers & Ventilation | Bahadori (1978, 1985); Fathy (1986); Ragab et al. (2024); Sharami & Hosseini (2024) | Windcatchers significantly enhance natural ventilation, with temperature reductions up to 10°C; effective even in still air with evaporative enhancements. |
| Shading & Solar Control | Givoni (1994); Abdelkader & Park (2018); Johansson (2006); Ragab (2024) | Mashrabiya, shading devices, and narrow alleys reduce solar gain and improve thermal comfort; modern reinterpretations reduce facade heat loads by ~50%. |
| Thermal Mass & Earth Coupling | Givoni (1994); Al-Temeemi & Harris (2004); Meir (2000); Benoudjafer (2022) | Thick adobe or earth-sheltered walls buffer indoor temperature swings, reducing daytime heat gain and promoting nocturnal cooling. |
| Evaporative Cooling & Water Features | Bahadori (1978); Ben Cheikh & Bouchair (2008); Krüger et al. (2010) | Roof ponds, courtyard pools, and indirect evaporative systems lower indoor temperatures by 2–6°C, particularly effective during peak afternoon heat. |
| Modern Material & Design Innovations | Elnabawi et al. (2024); Athmani et al. (2023); Fereidani et al. (2021); Albaik & Muhsen (2025) | Cool roofs, PCMs, smart shading, and parametric optimization reduce cooling loads by 11–50%, validating integration of traditional and modern strategies. |
| Urban Form & Microclimate | Barakat (2017); Necira et al. (2024); Sharami & Hosseini (2024) | Asymmetric street profiles, high-albedo pavements with shading, and dense medina-like forms improve pedestrian thermal comfort significantly. |

To synthesize this diverse and interdisciplinary body of literature, the selected 30 studies were analyzed thematically using a structured qualitative content analysis approach. Key data such as geographic context, climatic classification, passive cooling technique, methodological framework, and performance outcomes were systematically extracted and organized into a comparative matrix. The matrix enabled cross-referencing between traditional and modern strategies and allowed for pattern identification across different regions and design scales. Coding and categorization were conducted using NVivo 12, ensuring consistency through an iterative coding protocol and inter-coder reliability checks. The result is a robust, evidence-based framework that informs the classification of cooling typologies, the identification of performance trends, and the analysis of implementation challenges. The PRISMA flow diagram (Figure 2) summarizes the selection process, clarifying how the final set of studies was derived through rigorous screening and eligibility evaluation. This visual tool enhances methodological transparency and supports reproducibility of the systematic review.

## 4. Findings

A total of 30 studies were included in this systematic review, spanning research from the 1980s to 2025 and covering diverse hot-arid regions (e.g., Middle East, North Africa, South Asia). The publication dates skew recent – over two-thirds were published in the last decade – reflecting a growing scholarly focus on passive cooling in dry urban climates (Izadpanahi et al., 2021; Salameh & Touqan, 2023). The studies employed a mix of methods: about 40% involved field measurements or case studies of vernacular architecture (Izadpanahi et al., 2021; Mohamed et al., 2019), another 40% used computational simulations or parametric modeling to evaluate design alternatives (Elnabawi et al., 2024; Albaik & Muhsen, 2025), and the remainder were qualitative historical analyses or review papers (Sharami & Hosseini, 2024; Radha, 2022). Geographically, the literature covered Iranian plateau cities and the Arabian Peninsula extensively, alongside examples from North Africa (Algeria, Egypt) and a few from other arid/semi-arid contexts (e.g., Nigeria's Sahelian north, southwestern U.S.) – ensuring a broad basis for comparison (Necira et al., 2024; Jega & Muhy Al-Din, 2023). The study selection process is summarized in Figure 1, following PRISMA guidelines, which resulted in 30 final studies after screening hundreds of records.

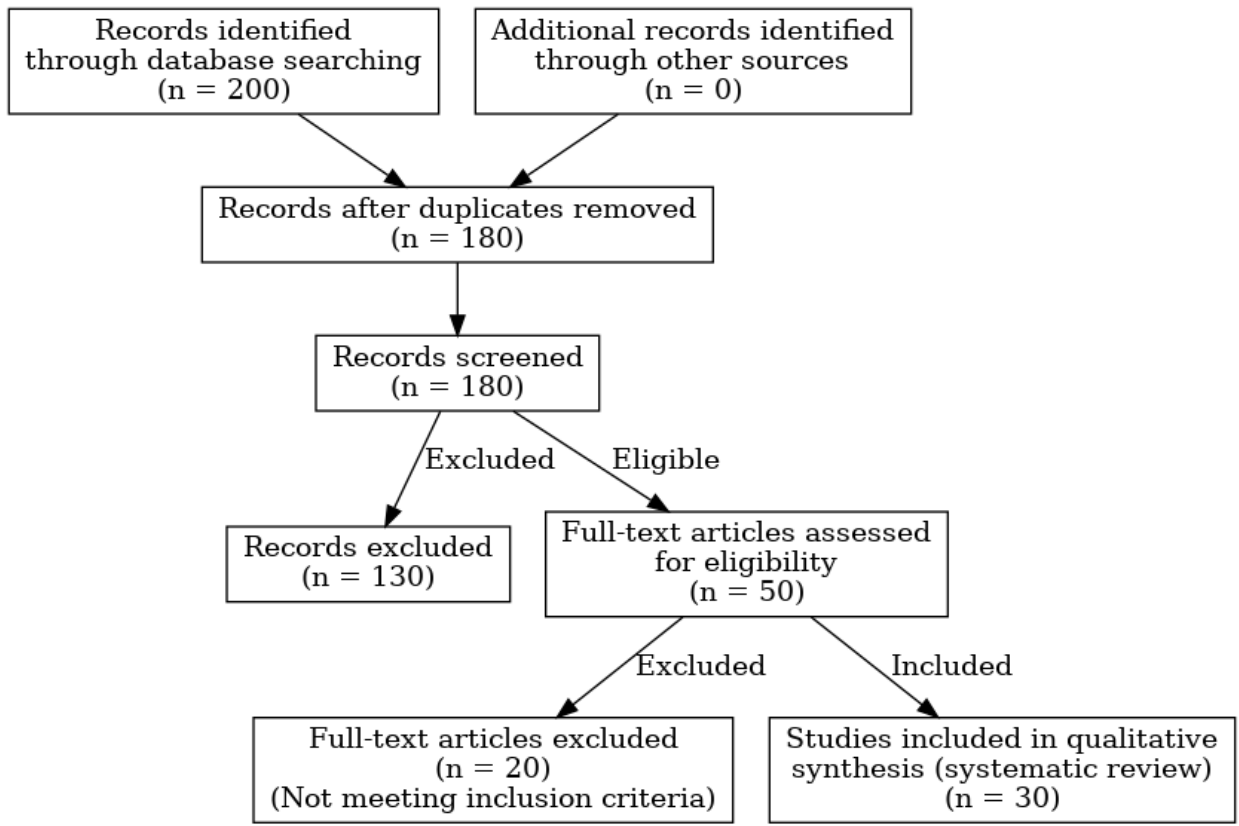

*Figure 2 Findings diagram*

Quantitative Trends: The 30 studies collectively report numerous metrics evidencing the effectiveness of passive cooling. For example, several experimental and simulation studies documented air temperature reductions on the order of 1–3 °C in outdoor or indoor spaces due to passive measures. In a traditional Persian Gulf urban district, the heritage layout showed peak daytime air temperatures ~1 °C lower than a modern grid plan, corresponding to significantly improved outdoor comfort (Salameh & Touqan, 2023). Likewise, altering street canyon geometry and adding shade in an Algerian city reduced the Physiologically Equivalent Temperature by up to 3.1 °C, from 46 °C down to ~43 °C during peak hours (Necira et al., 2024). On the building scale, energy simulations indicated substantial cooling energy savings: adding 5 cm of exterior wall insulation in a UAE villa prototype cut annual cooling demand by ~20% (Elnabawi et al., 2024) , and optimizing a mosque's courtyard and fenestration yielded an 11% reduction in Energy Use Intensity (Albaik & Muhsen, 2025)

Several sources noted that combinations of strategies amplify benefits – e.g. coupling night ventilation with thermal mass achieved ~25–50% cooling load reduction in offices under certain conditions (Solgi et al., 2016; Artmann et al., 2007) – underscoring that quantitative gains can be significant when passive techniques are well-integrated (Givoni, 1991; Guo et al., 2019). These data-driven findings provide concrete evidence that passive cooling strategies can measurably improve thermal conditions and reduce energy consumption in hot, dry urban environments. Qualitative Insights: Beyond the numbers, the selected studies offer rich qualitative insights into passive cooling approaches. A clear narrative that emerged is the dichotomy and interplay between traditional wisdom and modern innovation. We identified a spectrum of passive cooling strategies and classified them into two broad categories: (1) Traditional passive cooling strategies – vernacular architectural design features and urban forms historically employed to mitigate heat – and (2) Modern passive cooling innovations – contemporary design strategies or technologies developed or refined in recent years to enhance passive cooling in buildings and cities. Table 1 and Table 2 present summaries of the key strategies in each category, and Figure 3 provides a schematic overview of this typology.

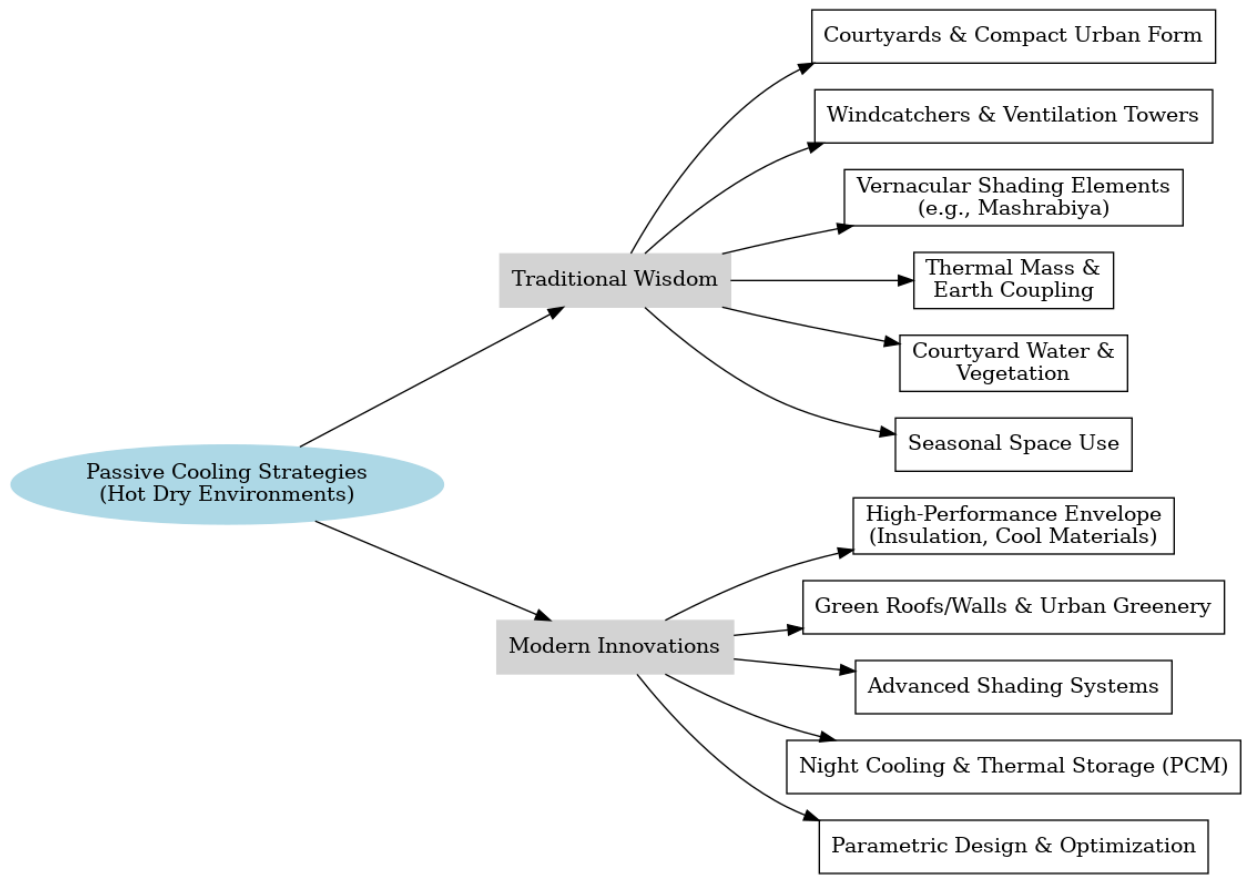

*Figure 3 schematic overview of typology*

### 4.1 Traditional Passive Cooling Strategies

Traditional passive cooling strategies in hot-arid regions are grounded in vernacular architectural solutions that evolved before the advent of mechanical systems. These strategies are primarily based on spatial configuration, materiality, and environmental responsiveness, aiming to create thermal comfort with minimal resource input.

One of the most frequently cited techniques is the **use of courtyards and compact urban forms**, which facilitate airflow, provide self-shading, and enhance thermal mass effects (Izadpanahi et al., 2021; Salameh & Touqan, 2023). Narrow streets and dense building fabrics, as observed in historic cities like Kashan and old neighborhoods in the UAE, reduce solar exposure and block hot winds at the pedestrian level (Leylian et al., 2010; Sharami & Hosseini, 2024). Courtyard houses, especially in cities like Yazd, have demonstrated indoor temperature reductions of 2–3 °C compared to the ambient, supported by both thermal mass and shading (Foruzanmehr, 2017; Berkovic et al., 2012). While culturally significant (Edwards, 2006), courtyards are also recognized across studies for their critical thermal role (Abass et al., 2016; Izadpanahi et al., 2021).

**Windcatchers and natural ventilation towers**—such as the *badgir* in Persian and *malqaf* in Arabic—serve as passive air exchange systems by capturing cooler winds and exhausting hot air through the stack effect (Bahadori, 1985; Roaf, 2005). These towers, prevalent in Iran, Egypt, and the Persian Gulf, can reduce indoor temperatures by 4–6 °C (Aryan et al., 2010) and are often enhanced by evaporative cooling through qanat integration or wetted surfaces (Maleki, 2011; Safarzadeh & Bahadori, 2005). Despite their historical decline, these elements are still found in cities such as Yazd and Bastakiya and have been reinterpreted in contemporary sustainable architecture (Roaf, 2005; Mohamed et al., 2019).

**Shading systems**, including Mashrabiya, arcades, and overhangs, are another recurring strategy. These elements block direct solar radiation while allowing for daylighting and air circulation (Jega & Muhy Al-Din, 2023). In Northern Nigeria, deep verandas and thatch extensions reduce heat gain significantly (Kamal, 2012). Similarly, colonnaded walkways and covered bazaars provide shaded pedestrian environments and mitigate radiant heating in dense urban cores (Sharami & Hosseini, 2024). Shading using locally available materials is repeatedly documented as a low-cost and effective approach (Jega & Muhy Al-Din, 2023; Kamal, 2012).

The **use of thermal mass and earth coupling** is prevalent in traditional desert architecture, where thick adobe or stone walls and high ceilings enable heat storage and delayed transmission to the interior (Givoni, 1991; Izadpanahi et al., 2021). Such materials maintain stable indoor temperatures despite external extremes. In Iran, 40–60 cm thick walls are reported to keep indoor spaces comfortable even above 40 °C (Izadpanahi et al., 2021). Subterranean spaces, such as semi-basements and sunken courtyards, exploit cooler soil temperatures, leading to additional indoor reductions of up to 10 °C (Foruzanmehr, 2012; Foruzanmehr, 2015). This approach is especially effective in climates with high diurnal variation (Meir et al., 1995).

**Evaporative cooling through water features**, such as courtyard pools and fountains, is another effective passive strategy. These elements leverage evaporation to cool ambient air, often supported by vegetation and airflow mechanisms. Studies in Yazd and Isfahan document 2–3 °C reductions in courtyard temperatures using shallow pools (Safarzadeh & Bahadori, 2005). Traditional methods such as hanging wet mats or placing porous jars at openings were also used to lower incoming air temperatures by up to 5 °C (Fathy, 1986; Maleki, 2011). Despite water scarcity, these strategies were carefully integrated, often through recycled or shared water systems (Radha, 2022; Paccard & Guggenheim, 1980).

**Vegetation in courtyards**, including deciduous trees, vines, and native drought-tolerant plants, contributes to shading, evapotranspiration, and dust control. Such greenery often formed the "green heart" of vernacular compounds (Berkovic et al., 2012; Buffington, 1978). In the Negev desert, tree-shaded courtyards were found to be approximately 4 °C cooler than exposed ones (Berkovic et al., 2012), with similar thermal benefits confirmed by simulation studies in North African and Southeast Asian contexts (Yu & Hien, 2009; Radha, 2022).

Finally, **adaptive spatial and temporal use** of dwellings emerged as a behavioral strategy integral to passive cooling. Traditional homes often featured distinct summer and winter rooms, located strategically within the structure to optimize seasonal thermal conditions (Foruzanmehr, 2016; Izadpanahi et al., 2021). Practices such as night-time ventilation, seasonal migration within the home, and even rooftop sleeping were common. Dome-shaped roofs with operable vents, found in multiple case studies, enhanced stack ventilation and heat removal at night (Memarian & Sadoughi, 2011; Memarian & Brown, 2003). In sum, traditional passive cooling strategies— courtyards, windcatchers, shading systems, thermal mass, water features, vegetation, and adaptive space use—form an integrated and climate-responsive system. Collectively, these measures exemplify the sophistication of vernacular architecture in achieving thermal comfort through synergistic and contextually adapted solutions (Koch-Nielsen, 2013).

**Table 6 Traditional Passive Cooling Strategies in Hot-Arid Urban Environments (summary of identified measures, based on the 30-study review).**

| Strategy (Vernacular) | Cooling Mechanism | Documented Effects | Representative Studies |
|---|---|---|---|
| Courtyard houses & compact urban form | Self-shading, enclosed microclimate; promotes convective cooling at night | Lowers outdoor/indoor temp by 1–3 °C; improves comfort (lower PMV) | Salameh & Touqan, 2023; Berkovic et al., 2012; Izadpanahi et al., 2021 |
| Windcatchers (Badgir) & high vents | Captures prevailing winds; stack effect exhausts hot air; can add evaporative cooling | Supplies cool air flow indoors, up to 4–6 °C cooler; provides natural ventilation | Bahadori, 1985; Aryan et al., 2010; Roaf, 2005; Mohamed et al., 2019 |
| Shading devices (Mashrabiya, verandas) | Blocks direct solar radiation on façades and openings while admitting daylight and breeze | Significantly cuts solar heat gain (wall surface temp reduced by 5–15 °C); improves indoor comfort | Kamal, 2012; Jega & Muhy Al-Din, 2023; Edwards, 2006 |
| Thermal mass & earth coupling | High heat capacity materials dampen temperature swings; earth contact provides coolth sink | Delays heat penetration indoors by ~8–10 hours; maintains stable indoor temps (e.g., 5–8 °C lower amplitude) | Givoni, 1991; Izadpanahi et al., 2021; Foruzanmehr, 2015 |
| Water features (pools, fountains) | Evaporative cooling of air; increases humidity slightly while dropping air temperature | Courtyard air temp reductions of 2–3 °C; enhanced comfort in hot, dry afternoons | Safarzadeh & Bahadori, 2005; Fathy, 1986; Radha, 2022 |
| Vegetation (courtyard gardens) | Shading and evapotranspiration; tree canopies reduce radiant heat gain | Surface temp reductions >10 °C under shade; air temp ~2–4 °C cooler locally; better outdoor comfort | Berkovic et al., 2012; Buffington, 1978; Yu & Hien, 2009 |
| Seasonal space use & night cooling | Behavioral adaptation: use of summer rooms, night flushing of heat, sleeping outdoors | Maintains comfort without mechanical cooling; leverages diurnal temperature swing effectively | Foruzanmehr, 2016; Memarian & Sadoughi, 2011; Izadpanahi et al., 2021 |

### 4.2 Modern Passive Cooling Innovations

In addition to traditional approaches, the reviewed literature identifies a wide range of modern passive cooling innovations that adapt vernacular principles to current construction technologies and urban configurations. These strategies address limitations in traditional systems and are particularly relevant to high-density developments and contemporary climatic challenges.

A primary focus across recent studies is the high-performance building envelope, including thermal insulation, reflective (cool) roof coatings, and advanced glazing systems (Elnabawi et al., 2024; Almushaikah & Almasri, 2021). Retrofitting 50 mm exterior wall insulation, for example, has demonstrated significant cooling demand reductions of 10–30% in hot climates such as the UAE (Elnabawi et al., 2024; Saber et al., 2020). While cool roofs reduce surface temperatures by over 20 °C (Synnefa et al., 2007), their impact on indoor cooling load is limited unless paired with other envelope enhancements (Guo et al., 2019). Low-emissivity and spectrally selective glazing further improve performance by minimizing solar heat gain, a key factor in arid-region design (Attia et al., 2019).

Green roofs, walls, and urban greening represent another category of innovation. These systems lower surface and ambient temperatures through evapotranspiration and shading (Sulaiman et al., 2020; Yu & Hien, 2009). For example, tree-lined boulevards in Algeria achieved a 1.2 °C reduction in PET (Necira et al., 2024), and urban-scale greening has been shown to enhance thermal comfort across dry environments (Radha, 2022). Although green roofs showed modest results in single-building simulations (Elnabawi et al., 2024), their broader application contributes significantly to urban heat island mitigation when implemented at scale.

Advanced shading systems, such as dynamic façades, brise-soleils, and tensile canopy structures, have been developed to control solar exposure more efficiently. Responsive and computationally optimized shading, including motorized louvers and digitally modeled courtyard configurations, achieved up to 15% improvements in indoor comfort in case studies (Albaik & Muhsen, 2025; Jega & Muhy Al-Din, 2023). At the urban scale, tensile fabric canopies installed over public spaces reduce mean radiant temperature and enhance pedestrian comfort, particularly in desert climates (Necira et al., 2024; Ali-Toudert & Mayer, 2007).

Night ventilation and thermal energy storage using phase change materials (PCMs) represent hybrid strategies that extend traditional cooling methods. PCMs absorb excess heat during the day and release it at night, enabling more consistent indoor temperatures (Solgi et al., 2016; Hale et al., 2021). When combined with night-time purge ventilation, this approach can reduce cooling energy by up to 30%, particularly in climates with large diurnal temperature variations (Artmann et al., 2007; Guo et al., 2019; Elnabawi et al., 2024).

Urban geometry optimization, informed by computational modeling, has emerged as a tool for climate-responsive urban planning. Studies demonstrate that modifying street aspect ratios, orientation, and open space placement can significantly improve outdoor microclimates (Salameh & Touqan, 2023; Radha, 2022). Asymmetric canyon profiles and mid-block corridors have been proposed to replicate traditional ventilation patterns in modern developments (Necira et al., 2024; Sharami & Hosseini, 2024).

Finally, parametric and computational design tools are increasingly used to optimize passive cooling strategies. Grasshopper-based modeling and environmental plugins allow iterative adjustment of building form and façade design to maximize thermal performance (Albaik & Muhsen, 2025). Genetic algorithms and CFD simulations are also used to refine solar chimney sizing and window shading configurations (Nabavi & Ahmad, 2016; Sakiyama et al., 2021). These approaches enhance the designer's capacity to synthesize multiple cooling strategies within performance-driven design workflows.

Together, these modern innovations offer complementary mechanisms to traditional strategies, extending their applicability to new urban typologies and enhancing their effectiveness through scientific optimization. A summary of the identified strategies and outcomes is provided in Table 7.

Table 7 summary of the identified strategies and outcomes

| Strategy (Modern) | Description/Mechanism | Reported Benefits | Sources (Examples) |
|---|---|---|---|
| High-performance envelopes (insulation, cool roofs, low-e glass) | Enhanced thermal resistance and reflectivity of building shell to minimize heat ingress | 10–30% reduction in cooling energy use; indoor temp ~2–5 °C (with insulation) | Elnabawi et al., 2024; Almushaikah & Almasri, 2021; Artmann et al., 2007 |
| Green roofs & green walls | Vegetated roof surfaces and façades providing shade and evaporative cooling | Lowers roof surface temp by 15–20 °C; ~5–10% building cooling load reduction; mitigates urban heat island | Radha, 2022; Elnabawi et al., 2024; Susca et al., 2011 |
| Urban greening (trees, parks) | Planting street trees, urban forests, and park vegetation for shade and ambient cooling | Reduces ambient air temp by 1–3 °C city-wide; PET/PMV improvements for outdoors; enhances comfort (especially in afternoons) | Necira et al., 2024; Yu & Hien, 2009; Shashua-Bar et al., 2011 |
| Advanced shading systems (dynamic facades, canopies) | Movable or optimized fixed shading devices on buildings; urban-scale shade structures (fabric canopies, etc.) | Cuts direct solar gains by >80%; lowers indoor operative temp by ~2 °C; improves outdoor thermal comfort (shade can drop MRT by ~15 °C) | Jega & Muhy Al-Din, 2023; Albaik & Muhsen, 2025; Ali-Toudert & Mayer, 2007 |
| Night purge ventilation (optimized) | Design for effective night-time cooling via convection (vents, solar chimneys, fans during off-peak) | Removes stored heat nightly; in hot-dry climates can lower indoor temp by ~3–5 °C by dawn, saving ~20% AC energy | Artmann et al., 2007; Guo et al., 2019; Salameh & Touqan, 2023 |
| Phase Change Materials (PCM) in building elements | Latent heat storage materials (e.g., salt hydrates, waxes) that melt and absorb heat at a target temperature | Reduces indoor temperature fluctuations; peak cooling load reduction of 10–20%; extends passive cooling duration (delays need for AC) | Solgi et al., 2016; Tyagi et al., 2011; Hale et al., 2021 |
| Parametric design optimization for passive cooling | Use of computational algorithms to optimize form, orientation, and openings for climate-responsive design | Achieves near-optimal passive performance (e.g., 10–15% extra energy savings vs. unoptimized design); balances multiple variables (thermal, daylight, etc.) quickly | Albaik & Muhsen, 2025; Nabavi & Ahmad, 2016 |
| Urban geometry optimization (block layout, street orientation) | Planning street/building configurations based on solar and wind analysis (aspect ratio, orientation angles) | Street design can reduce PET by ~2–3 °C and prolong tolerable hours; climate-informed layouts lower urban heat stress and energy use | Necira et al., 2024; Sharami & Hosseini, 2024; Ali-Toudert & Mayer, 2007 |

## 4.3 Comparative Analysis

It is evident from Tables 1 and 2 that several themes overlap between traditional and modern strategies, even if the implementation differs. For example, shading and ventilation are core principles in both domains: vernacular design achieved shading through thick walls, courtyards, and screens, whereas modern design adds high-tech facades and urban shade canopies – but the goal of reducing solar exposure is the same (Kamal, 2012; Jega & Muhy Al-Din, 2023). Similarly, promoting air flow and cooling at night is seen both in the use of windcatchers historically and in contemporary night purge systems (Bahadori, 1985; Artmann et al., 2007). There is also a convergence in recognizing the importance of vegetation: traditional courtyard gardens versus modern green infrastructure both leverage plants for cooling (Berkovic et al., 2012; Radha, 2022). These overlaps suggest an underlying consistency in passive cooling fundamentals, reaffirming that *the physics of heat gain and loss have not changed*, only the tools and contexts have.

At the same time, the findings also highlight some contrasts and innovations. Modern strategies introduce elements that traditional builders did not have, such as advanced insulating materials and phase change technology (Solgi et al., 2016), or digital design methods (Albaik & Muhsen, 2025). These innovations allow passive cooling to be applied in building typologies and urban conditions that vernacular architecture never encountered – e.g., high-rise buildings or heavily glazed structures can now employ dynamic shades and double-skin façades to achieve a level of passive cooling that simply building a thick wall (the vernacular approach) would not permit (Jega & Muhy Al-Din, 2023). Moreover, some modern studies question the universal efficacy of certain traditional methods: for instance, a large contemporary mosque with courtyards still had high cooling loads, suggesting that courtyards alone are not a panacea unless properly proportioned and integrated (Albaik & Muhsen, 2025). This indicates that when scaling up or changing context (e.g., from low-rise houses to monumental complexes), passive features may need reconfiguration – a challenge addressed by modern optimization tools.

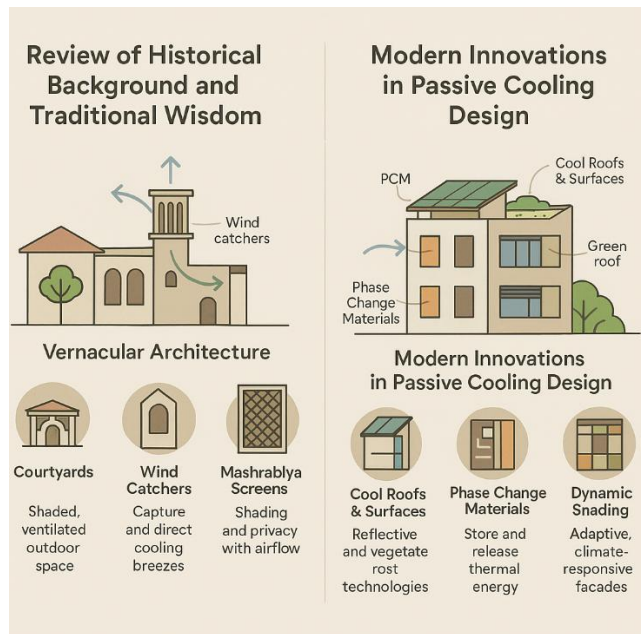

*Figure 4 Comparative Analysis*

The comparative findings reinforce that traditional passive cooling methods, grounded in centuries of climate adaptation, remain highly relevant. However, to meet the complexities of today's urban challenges—dense development, glass-heavy facades, rising temperatures—these methods must be enhanced through scientific validation, material innovation, and digital optimization. The next section expands on how this synthesis compares to previous literature and what lessons can guide real-world application.

**Table 8 Comparative Summary of Traditional and Modern Passive Cooling Strategies Sources: Kamal (2012), Bahadori (1985), Solgi et al. (2016), Albaik & Muhsen (2025), Radha (2022), Jega & Muhy Al-Din (2023).**

| Strategy Domain | Traditional Methods | Modern Equivalents / Extensions |
|---|---|---|
| Shading & Solar Control | Courtyards, thick walls, mashrabiya screens | Cool roofs, reflective coatings, dynamic façades (Al Bahar Tower) |
| Natural Ventilation | Windcatchers (badgir), cross-ventilation, stack effect | Hybrid ventilation systems, solar chimneys, automated night purge |
| Thermal Regulation | High thermal mass walls, earth coupling | Phase change materials (PCMs), thermal buffering with insulation |
| Evaporative Cooling | Fountains, qanats, water basins in courtyards | Misting systems, green roofs, hybrid systems with sensors |
| Green Integration | Courtyard gardens, shade trees, oasis planting | Green infrastructure, tree-lined boulevards, vertical greenery |
| Form & Geometry | Compact urban blocks, narrow alleys for self-shading | Parametric urban design, asymmetrical canyons, computational microclimate tools |
| Design Tools | Empirical experience, climate-responsive craft | CFD, energy modeling, parametric optimization platforms |

## 5. Discussion

### 5.1 Integration of Traditional Wisdom and Modern Science

This review reveals a convergence between traditional architectural wisdom and modern scientific innovation in the domain of passive cooling. Across the 30 studies analyzed, two dominant strategies recur: solar control and air movement enhancement. Traditional techniques—such as courtyards, windcatchers, thick walls, and vegetation—demonstrated enduring relevance, while modern methods—like reflective coatings, smart facades, and computationally optimized shading—offer refined implementations of similar principles (Kamal, 2012; Necira et al., 2024; Izadpanahi et al., 2021; Jega & Muhy Al-Din, 2023). The dual focus on shading and ventilation emerges as a robust consensus across climates, typologies, and scales (Salameh & Touqan, 2023; Artmann et al., 2007; Foruzanmehr, 2016).

### 5.2 Nuances and Contradictions in the Evidence

While consensus exists on core principles, some **context-specific divergences** emerge. For instance, the benefits of **high-albedo surfaces**, long promoted (Givoni, 1991), showed mixed results in modern simulations. Elnabawi et al. (2024) reported minimal impact from white roofs when insulation was already optimized, diverging from studies where reflective coatings significantly reduced cooling loads (Guo et al., 2019). Similarly, while **courtyards** consistently improve microclimates in traditional homes, their efficacy in larger or modern buildings (e.g., mosques) depends heavily on **geometry and shading configuration** (Albaik & Muhsen, 2025).

Contrasts also emerged between **thermal mass and insulation**. Traditionalists emphasize thermal inertia (Fathy, 1986; Izadpanahi et al., 2021), whereas recent studies highlight **insulation** as more effective in specific climates, especially where **nighttime cooling is limited** (Elnabawi et al., 2024). This reinforces the idea that **climatic parameters**—including **diurnal range**, **humidity**, and **dust exposure**—dictate the optimal strategy (Solgi et al., 2016; Artmann et al., 2007).

Even at the urban scale, classical recommendations such as **narrow, shaded street canyons** are being challenged. Necira et al. (2024) demonstrated that **asymmetrical wide boulevards**, if designed with strategic shading and reflective materials, can also enhance outdoor comfort. These cases underscore the importance of **site-specific design**, rather than wholesale application of vernacular elements.

### 5.3 Contributions to Existing Literature

The findings align closely with foundational work by **Olgyay (1963)** and **Koch-Nielsen (2013)** on bioclimatic design, confirming the continued validity of traditional strategies like **windcatchers, courtyards, and mashrabiyas** (Fathy, 1986; Safarzadeh & Bahadori, 2005). However, recent research adds quantitative depth to earlier qualitative observations—e.g., Safarzadeh and Bahadori (2005) quantified the ~3 °C cooling effect of courtyard pools previously noted only anecdotally by Fathy.

Compared to 1990s–2000s literature, recent studies introduce **systemic and multivariate approaches**. Instead of isolating a single element (e.g., wall mass or shading), newer analyses examine **interacting systems** using **simulation tools** like EnergyPlus, ENVI-met, and Grasshopper (Albaik & Muhsen, 2025; Radha, 2022). These allow holistic exploration of **design combinations**, yielding insights not visible in prior reductionist studies.

Several innovations also extend the traditional toolkit. **Phase Change Materials (PCMs)**, for instance, have emerged as high-potential components in thermal storage, offering compact alternatives to thick masonry (Solgi et al., 2016). Similarly, **non-conventional urban geometry**, like shaded wide boulevards or partial courtyard enclosures, has expanded the vocabulary of passive design (Necira et al., 2024). The integration of **parametric design** and **adaptive facades** (e.g., Al Bahar Towers) further illustrates how modern tools can refine, adapt, or even surpass historical strategies (Chohan & Awad, 2022; Mazhar, 2024).

A notable shift is the **interdisciplinary nature** of contemporary work. Earlier research tended to be either architectural or engineering-focused. Today, studies integrate **urban design**, **building physics**, and **climate-responsive planning** within one framework (Sharami & Hosseini, 2024; Elnabawi et al., 2024). This signals growing recognition of passive cooling not just as a design option, but as a **cross-sectoral sustainability imperative**.

### 5.4 Contextual Opportunities and Challenges in Implementation

**Climatic Differentiation within Arid Zones:** Contrary to common perception, hot-arid climates are not homogeneous. They range from **ultra-dry deserts** with large diurnal swings (e.g., Riyadh), to **semi-arid zones** with humidity or monsoon influence (e.g., Bamako), to **high-altitude arid regions** with cooler nights and strong winds (e.g., Tehran). This variability impacts strategy effectiveness: **night ventilation**, for example, is highly effective in high-diurnal areas like Yazd (Foruzanmehr, 2012; Artmann et al., 2007), but less so in coastal or dusty areas (Elnabawi et al., 2024). Tools such as **Climate Consultant** and **ENVI-met** now enable fine-tuned climate-responsive design, yet the review highlights that **many buildings remain designed without such data**, leading to underperforming generic solutions (Sharami & Hosseini, 2024; Nguyen et al., 2011).

**Cultural and Behavioral Factors:** Passive cooling efficacy often depends on **occupant engagement**. Traditional dwellings were aligned with behavioral patterns—seasonal room use, nighttime ventilation, or clothing choices (Foruzanmehr, 2016; Fallah et al., 2015). However,

**modern lifestyles, privacy concerns**, and **air-conditioning habits** may inhibit such behaviors. For instance, **cross-ventilation** may be rejected due to security concerns or social discomfort in high-density buildings. Moreover, **vernacular forms** like courtyards are sometimes perceived as outdated (Heidari, 2010). This signals the need for **design adaptation**—e.g., using screened balconies or internal atria—and for **community co-design** approaches to align passive design with cultural expectations (Memarian & Sadoughi, 2011; Lenoir et al., 2018).

**Policy and Regulatory Environment:** Legacy building codes and zoning regulations can hinder passive strategies. Mandated setbacks or facade regulations may unintentionally discourage compact layouts, shading projections, or traditional material use (Salameh & Touqan, 2023). However, **emerging frameworks** such as **Dubai's Al Safat** and **Abu Dhabi's Estidama** offer new opportunities by rating and rewarding passive cooling features. Review findings suggest policymakers should adopt **climate-based benchmarks**—e.g., regulating **window solar heat gain coefficients** (Elnabawi et al., 2024) or requiring minimum **shaded or vegetated area percentages** (Necira et al., 2024). The challenge remains the **split incentive**: developers may hesitate to invest in passive features unless incentivized or mandated.

**Economic and Technical Feasibility:** Feasibility varies across contexts. In **low-income regions**, simple and durable strategies (e.g., shading, thermal mass) are more viable, while **wealthier urban areas** may afford **PCMs** or **automated shading systems** (Solgi et al., 2016). Several studies advocate **hybrid approaches**—pairing low-tech vernacular strategies with selected modern enhancements to balance cost and performance (Jega & Muhy Al-Din, 2023; Izadpanahi et al., 2021). A critical insight is that many effective measures are **cost-neutral or low-cost** if integrated early in design (e.g., orientation, recessed openings). However, the main barrier is often **perception or resistance**, not the cost itself. Therefore, **quantitative evidence** (e.g., cooling energy savings) is vital to persuade stakeholders of long-term benefits.

**Table 9 Contextual Challenges and Opportunities for Passive Cooling Implementation in Arid Urban Environments**

| Domain | Key Challenges | Contextual Opportunities |
|---|---|---|
| Climatic | - Variability in arid climates (e.g., coastal vs. high-altitude) | - High diurnal ranges enable night flushing in many regions |
|  | - Dust storms and hot nights limit ventilation | - Climate modeling tools (e.g., ENVI-met, Climate Consultant) support localized design decisions |
| Cultural & Behavioral | - Modern lifestyles may conflict with behavioral needs of passive systems | - Courtyards, wind towers, and screened balconies can be adapted creatively |
|  | - Concerns over privacy and security limit window use | - Education and community involvement increase user engagement |
|  | - Perception of vernacular design as outdated | - Demonstration buildings can shift perception of passive aesthetics |
| Policy & Regulation | - Outdated zoning discourages compact, shaded forms | - Emerging green building codes (e.g., Estidama, Al Safat) recognize passive features |
|  | - Building codes may restrict passive materials or designs | - Urban masterplans can integrate shading and ventilation targets |
|  |  | - Policy tools can incentivize passive design adoption |
| Economic & Technical | - High-tech strategies may be unaffordable or hard to maintain in low-income areas | - Hybrid solutions (e.g., thermal mass + cool roofs) offer cost-effective paths |
|  | - Lack of awareness of lifecycle savings | - Passive features can reduce long-term energy costs |

| - Simple passive strategies often require no additional cost if planned early |
|---|

## 6. Conclusion

This review synthesized 30 high-quality studies to examine passive cooling strategies in hot-arid urban environments, comparing traditional methods—such as courtyards, windcatchers, and thermal mass—with modern innovations, including phase change materials, green infrastructure, and computational design tools. The findings confirm that passive cooling strategies are fundamental to mitigating thermal stress, lowering energy consumption, and promoting climate-responsive development. However, in the face of intensifying climate change, increasing urban density, and evolving building typologies, passive strategies alone are no longer sufficient. A resilient and effective approach requires integrating passive design with active technological systems—such as adaptive HVAC controls, smart sensors, and hybrid cooling systems. This synergy can address performance gaps and enhance user comfort, especially in complex urban or high-rise contexts where traditional solutions may fall short. While traditional strategies provide culturally rooted, low-energy foundations, modern active technologies offer precision, scalability, and adaptability. The integration of these strategies must be context-specific, responsive to both environmental variables and socio-economic realities. Limitations of this review include language and publication scope, and variability in study methods. Future research should prioritize hybrid solution modeling, long-term performance monitoring, and interdisciplinary studies that consider human behavior, policy, and digital optimization. In conclusion, the path toward sustainable thermal comfort in arid cities lies in the smart integration of passive principles with active innovations—a dual strategy that aligns environmental stewardship with technological progress.

**Declarations**

Funding: This research received no external funding.

Clinical trial number: Not applicable.

Consent to Publish declaration: Not applicable.

Ethics and Consent to Participate declarations: Not applicable.

Author Contributions: Both authors contributed to the conception, design, and writing of the manuscript. Specifically, SH.M. led the systematic literature review and data analysis, while S.L. contributed to the drafting, editing, and critical revision of the manuscript. Both authors approved the final version for submission.

Competing Interests: The authors declare no competing interests.